\documentstyle[preprint,aps]{revtex}
\input epsf.tex
\def \half { {1 \over 2}}
\def \d {\delta}

\def \ds {\not \! \partial}

\def\overlay#1#2{\setbox0=\hbox{#1}\setbox1=\hbox to \wd0{\hss
#2\hss}#1\hskip -2\wd0\copy1}
\def\lsim{\mathrel{\rlap{\lower4pt\hbox{\hskip1pt$\sim$}}
    \raise1pt\hbox{$<$}}}         
\def\gsim{\mathrel{\rlap{\lower4pt\hbox{\hskip1pt$\sim$}}
    \raise1pt\hbox{$>$}}}         
\def\beq{\begin{equation}}
\def\eeq{\end{equation}}
\def\bea{\begin{eqnarray}}
\def\eea{\end{eqnarray}}

\def\eps{\epsilon}

\begin{document}
\preprint{\hfill {PM 95-36\\}}
\title{Variational solution of the Gross-Neveu model at finite temperature
in the large $N$ limit.}
\author{B. Bellet, P. Garcia, F. Geniet and M.B. Pinto \\
Laboratoire de Physique Math\'ematique, Universit\'e de Montpellier II \\
CNRS-URA 768, 34095 Montpellier Cedex 05, France}
\maketitle

\begin{abstract}
We use an improved nonperturbative variational method to
investigate the phase transition of the Gross-Neveu model.
It is shown that the variational procedure can be generalized
to the finite temperature case. The large $N$ result for the phase transition
is correctly reproduced.
\end{abstract}
\newpage
\section{Introduction}
\label{sec:intro}

A survey of the literature concerning analytical nonperturbative methods shows
a growing interest in variational methods such as the gaussian approximation
\cite {pm} and
the optimized linear $\d$ expansion \cite {suk}. Recently, an alternative
variational method \cite {an} has been improved and successfully tested in the
Gross-Neveu (GN) model \cite {gn} both in the large \cite {fred} and finite
\cite {fred1} $N$ limits. Basically this new approximation differs from the
existing optimized linear $\d$ expansion in two aspects: it provides a clear
way of carrying on the renormalization program of the theory and improves the
calculation with
a simple but powerful analytical method which allows to attain arbitrarily
large and even infinite order of perturbation in the parameter $\d$.
Our goal is to show that this calculational scheme, especially the variational
procedure, still works when external parameters such as the temperature are
considered. From the renormalization point of view the temperature does not
create any new difficulties and the elimination of divergencies can be carried
out exactly as at zero temperature \cite {fred}. However, previous experience
with the related $\d$ expansion shows that the application of variational
methods to the study of phase transitions is not necessarily straightforward
\cite {mar}.

Here, our aim is to check the ability of the variational procedure used in
Ref. \cite {fred} in generating accurate results for the phase transition,
especially near the critical temperature. In this first finite temperature
application we restrict ourselves to the large $N$ limit of the GN model where
the calculational scheme can be set up more clearly. For comparison we consider
the
results of Refs. \cite {gnft} and \cite {hay} as being ``exact" in the large
$N$ limit, despite their disagreement with existing theorems \cite {landau} for
symmetry breaking/restoration in 1+1 dimensions. This interesting problem,
which has been studied by several authors using different models [11-15],
will be explicitly treated in an extension of the present work \cite {beno}.

In the next section we review the usual
large $N$ result for the temperature dependent fermionic mass and perform
the variational calculation to lowest orders. We shall see that, as in the zero
temperature case,
the variational procedure fails. In Section III we follow Ref. \cite {fred} to
rectify the situation
by performing the variational calculation to all orders. When this is done
the large $N$ result is completely reproduced. The conclusions and future
perspectives are presented in section IV.

\section{The variational calculation to lowest orders }
The variational calculation starts with the addition of an arbitrary bare mass
($m_0$) to the original  massless  Gross-Neveu \cite {gn}

\beq
{\cal L}= i \sum_{i=1}^N \bar \psi_i \ds \psi_i + m_0 \sum_{i=1}^N \bar \psi_i
\psi_i + \frac{g_0^2}{2} \left ( \sum_{i=1}^N \bar \psi_i \psi_i \right )^2
\label {gro}
\eeq
where $g_0$ is the bare coupling constant (in the following we shall suppress
the summation over the index $i$). The relation to the linear $\d$
expansion and other variational methods becomes clear by performing the
substitutions
\beq
m_0 \rightarrow m_0(1-\d) \label {sub1}\;\;\;\;,
\eeq
\beq
g_0^2 \rightarrow \d g_0^2 \label {sub2}\,\,\,.
\eeq
These will be done at a later stage in order to avoid the explicit evaluation
of Feynman graphs which differ only by $\d m_0$ insertions.
To perform finite temperature calculations in the imaginary time formalism
one does the following substitutions~\cite{ftemp}

\beq
\int \frac {dp_0}{2\pi} \rightarrow  i T \sum_n \;\;\;,
\eeq
\beq
p_0 \rightarrow i \omega_n \;\;\;,
\eeq
where for fermions

\beq
\omega_n = T (2n + 1) \pi\;\;\;.
\eeq
The sum over Matsubara's frequencies can be performed with

\beq
T \sum_n \ln (\omega_n^2 + m_0^2) = E + 2 T \ln [ 1 + \exp(-E/T)] +c
\,\,\,\,,
\eeq
where c is a $E$-independent constant. In 1+1 dimensions

\beq
E=(p_1^2 + m_0^2)^{\half}\;\;\;.
\eeq
The remaining space integral is evaluated in $1-\eps$ dimensions using
conventional dimensional regularization techniques.

In the large $N$ limit a perturbative calculation of the fermionic mass
($M_F$) to O($g_0^4$) yields
\beq
M_F^{(2)} = m_F^{(0)} + g_0^2 m_F^{(1)} + g_0^4 m_F^{(2)} + {\rm O}(g_0^6)
\;\;\;,
\eeq
where $m_F^{(0)} = m_0$,
\beq
m_F^{(1)} =  \frac {N}{2\pi} m_0^{1-\eps} \left [  ( 4 \pi)^{\frac{\eps}{2} }
 \Gamma (\eps/2) -  4 I_1^T(y_0)\right ] \;\;\;,
\eeq
and

\begin{eqnarray}
 m_F^{(2)} & = &  \nonumber  \frac {N^2}{4\pi^2}  m_0^{1-2\eps}
                \left\{ (4 \pi)^{\eps}  \Gamma^2 (\eps/2)(1 - \eps) \right. \\
           &  & \nonumber
                + 4 (4 \pi)^{\frac {\eps}{2}}  \Gamma(\eps/2)
                 \left [ (3\eps - 2) I_1^T(y_0)
                 +  (m_0/T)^2 (1-\eps)(I_2^T(y_0)+
                  I_3^T(y_0) ) \right ]    \\
           &  & \left.
                 + 16 I_1^T(y_0)\left [ I_1^T(y_0) - (m_0/T)^2
                  (I_2^T(y_0)+
                  I_3^T(y_0) ) \right ] \right \} \label {o2} \;\;\;.
\end{eqnarray}
The temperature dependent integrals are
\beq
I_1^T(y_0)= \frac{\pi^{\frac{1-\eps}{2}}}{\Gamma(1/2 -\eps/2)} (2\pi
y_0)^{\eps}   \int_0^{\infty} dx \frac {x^{-\eps}}{(x^2+y_0^2)^{\half}
[1+\exp(x^2+y_0^2)^{\half}]} \;\;\;,
\eeq

\beq
I_2^T(y_0)=\frac{\pi^{\frac{1-\eps}{2}}}{\Gamma(1/2 -\eps/2)} (2\pi y_0)^{\eps}
  \int_0^{\infty} dx \frac {x^{-\eps}}{(x^2+y_0^2)^{\frac{3}{2}}
[1+\exp(x^2+y_0^2)^{\half}]}  \;\;\;,
\eeq
and
\beq
I_3^T(y_0)=\frac{\pi^{\frac{1-\eps}{2}}}{\Gamma(1/2 -\eps/2)} (2\pi y_0)^{\eps}
  \int_0^{\infty} dx \frac {x^{-\eps}\exp(x^2+y_0^2)^{\half}}{(x^2+y_0^2)
[1+\exp(x^2+y_0^2)^{\half}]^2}  \;\;\;,
\eeq
with $y_0= m_0/T$ and $x=p_1/T$. These integrals are related to each other via
\beq
- -2 y_0^{\eps} \frac {\partial (y_0^{-\eps}I_1^T)}{\partial y_0^2} = I_2^T +
I_3^T
\;\;\;.
\eeq

The nonperturbative large $N$ calculation consists in the evaluation of all
cactus diagrams which can be summed up as
\beq
M_F = m_0 \left \{1- \frac {N}{2\pi} g_0^2 M_F^{-\eps} \left [
(4 \pi)^{\frac{\eps}{2} }
 \Gamma (\eps/2) -  4 I_1^T(y_F) \right ] \right \}^{-1} \label {mass}\;\;\;,
\eeq
where $y_F = M_F/T$.
The bare and renormalized parameters are related via
\beq
m_0 =  Z_m m  \label {zm}\;\;\;,
\eeq
and
\beq
 g_0^2 =  Z_g  g^2 \mu^{\eps} \label {zg}\;\;\;,
\eeq
where $\mu$ is the arbitrary scale introduced by dimensional regularization.
The renormalization constants are

\beq
Z_m = Z_g= \left [ 1 + \frac{g^2 N}{\pi \eps} \right ]^{-1} \;\;\;.
\eeq
Substituting Eqs.~(\ref{zm}) and (\ref{zg}) into Eq.~(\ref{mass}) yields
the finite expression for the dimensionless quantity $
{M_F}/{\Lambda_{\overline {MS}}}$

\beq
\frac {M_F}{\Lambda_{\overline {MS}}} =
 \frac {m}{\Lambda_{\overline{ MS}}}
\left \{1 + \frac{N}{\pi} g^2  \left [ \ln \left ( \frac {M_F}{\bar \mu}
\right ) + 2 I_1^T(y_F) \right ] \right \}^{-1} \label {dimless}\;\;\;\;,
\eeq
where $\Lambda_{\overline {MS}} = \bar \mu \exp[-\pi/(N g^2)]$ and
 $\bar \mu = \mu (4\pi)^{\half}\exp(-\gamma_E/2)$ with $\gamma_E = 0.577215$.
Equation (\ref{dimless}) satisfies the RG equation

\beq
\mu \frac { d}{d \mu} M_F = \left ( \mu \frac{\partial}{\partial \mu}
+ g \beta(g) \frac{\partial}{\partial g} - m \gamma_m(g)
\frac{\partial}{\partial
 m} \right ) M_F = 0 \;\;\;,
\eeq
where

\beq
\beta(g) = - \frac {N}{2 \pi}g^2 \;\;\;{\rm and}\;\;\;
\gamma_m(g) = \frac {N}{\pi} g^2 \;\;\;.
\eeq
Then, in the chiral limit ($m=0$), the finite temperature mass
gap equation is given by
\beq
{M_F(T)}=  {M_F(0)}
\exp[-2 I_1^T(y_F)] \label {gap}\,\,\,,
\eeq
where $M_F(0) = \Lambda_{\overline {MS}}$ is the renormalized fermion mass,
generated through dynamical symmetry breaking at $T=0$. This is just the  large
$N$
result which reproduces the phase diagram for the fermion mass \cite {gnft}.

The variational calculation starts with the substitutions Eqs.~(\ref{sub1}) and
(\ref{sub2}) . With this procedure Eq.~(\ref{gro}) interpolates between the
original massless GN model (at $\d=1$) and a massive free theory (at $\d=0$).
The perturbative calculation is now done in powers of the bookkeeping parameter
$\d$ and extremized with respect to $m_0$
at $\d=1$ (Principle of Minimal Sensitivity \cite {pms}).
Starting with O($\d$) one has at $\d=1$
\beq
M_F^{(1)}(m_0)=m_F^{(0)}+g_0^2 m_F^{(1)} - m_0 \frac{\partial m_F^{(0)}}
{\partial m_0} = g_0^2 m_F^{(1)}\;\;\;,
\eeq
which has no nontrivial extremum in $m_0$ at $T=0$ nor at finite temperatures.
At $\d=1$ the O($\d^2$) fermionic mass is given by
\beq
M_F^{(2)}(m_0)= g_0^2m_F^{(1)} - g_0^2 m_0 \frac{\partial m_F^{(1)}}{\partial
m_0} + g_0^4 m_F^{(2)} \;\;\; \label {m2}.
\eeq
The extremization with respect to $m_0$ in the limit $\eps \rightarrow 0$ does
not give any useful information since the term $g_0^4 m_F^{(2)}$, which has no
nontrivial extremum, dominates. This behaviour, which persists to higher
orders, has also been noted at zero temperature \cite {fred}
and we find that the situation does not change at finite temperatures.

\section {The variational calculation to all orders.}
In this section we shall follow
Ref. \cite{fred} to perform an all order variational calculation  in the
large $N$ limit eliminating the optimization problem encountered in the
previous section. The general philosophy within variational methods is to start
with a trial value which is expected to be reasonably close to the true value
of the
physical parameters. In our case this means that we can start by formulating a
nonperturbative ansatz which already resums a good part of the RG behaviour of
the fermionic mass before launching into the actual variational calculation. Of
course this is a rather easy task within the large $N$ limit, where the
exact answer Eq.~(\ref{mass}) constitutes the natural choice.
Performing the substitutions Eqs.~(\ref{sub1}) and (\ref{sub2}) in
Eq.~(\ref{mass}) we get
\beq
M_F(f)= \frac {m_0 (1-\d)}{f(\d)}\;\;\;\;,
\eeq
where we have defined
\beq
f(\d)= 1- \frac {N}{2\pi} \d g_0^2 m_0^{-\eps}(1-\d)^{-\eps} f^{\eps} \left [
(4 \pi)^{\frac{\eps}{2} }
 \Gamma (\eps/2) -  4 I_1^T(y_F^{\prime}) \right ] \;\;\;,
\eeq
with $y_F^{\prime}=M_F(f)/T$.
It is now possible to perform an expansion in powers of $\d$
to order-$n$ around the free theory ($\d=0$). Using contour integration
one obtains $M_F$ to $n^{\rm th}$ order of perturbation theory

\beq
M_F^{(n)}(m_0)=\frac{1}{2 \pi i} \oint
dz(\frac{1}{z}+\frac{\d}{z^2}+...+\frac{\d^{n}}{z^{n+1}})
\frac{m_0(1-z)}{f(z)}\;\;\;,
\eeq
which, at $\d=1$, gives
\beq
M_F^{(n)}(m_0)=\frac{1}{2 \pi i} \oint \frac {dz}{z^{n+1}}
\frac{m_0}{f(z)}\;\;\;.
\eeq
Once $Z_m$ and $Z_g$ are applied to the bare $m_0$ and $g_0^2$ one gets the
finite expression for the dimensionless quantity $M_F/
\Lambda_{\overline {MS}}$

\beq
\frac {M_F^{(n)}(m)}{\Lambda_{\overline {MS}}} =
\frac{1}{2\pi i} \oint \frac{dz}{z^{n+1}} \frac {m}{\Lambda_{\overline{ MS}}}
\left \{1 + \frac{N}{\pi} g^2 z \left [ \ln \left ( \frac {M_F(f)}{\bar \mu}
\right ) + 2 I_1^T(y_F^{\prime}) \right ] \right \}^{-1} \;\;\;.
\eeq
As noted in the case of the anharmonic oscillator \cite {phi}, it is possible
to extract more structure from the limit of infinite order by rescalling $m$
with the order $n$. After distortion of the contour it is clear that only the
vicinity of
$z=1$ survives in the limit $n \rightarrow \infty$, which can be analyzed by
changing variables
\beq
1- z = \frac {v}{n} \;\;\;.
\eeq
Rescalling $m$ by introducing $m^{ \prime}=m/n$ we get, in the
$n\rightarrow \infty$ limit
\beq
\frac{M_F(m^{ \prime \prime})}{\Lambda_{\overline{MS}}}= \frac {1}{2\pi i}
\oint
\frac {dv e^v m^{ \prime \prime}}{K(v)} \label {oint}\;\;\;,
\eeq
where the integration runs counterclockwise around the negative real axis. The
function $K(v)$ is given by
\beq
K(v) = \ln \left ( \frac { m^{\prime \prime} v}{K(v)} \right ) +2 I_1^T \left (
\frac { m^{\prime \prime} v}{t K(v)} \right ) \label {kv} \;\;\;,
\eeq
where
\beq
m^{ \prime \prime }= \frac {m^{ \prime}}{\Lambda_{\overline {MS}}} \left (
\frac
{N g^2}{\pi} \right )^{-1} \;\;\;,
\eeq
with $t = T/ \Lambda_{\overline{MS}}$.
Equations (\ref{oint}) and (\ref {kv}), which summarize our variational
approach, should be understood as follows: for a given variational parameter
$m^{\prime \prime}$, and
a given temperature $T$, Eq.~(\ref {kv}) enables one to determine $K(v)$ self
consistently. The variational result is then given in an explicitly RG
invariant way by extremizing Eq.~(\ref {oint}) with respect to
$ m^{\prime \prime}$. In general this program has to be achieved numerically.
However, before doing that one can use the fact that the large $N$ limit is
free from infra red divergences to perform an analytical exploitation of the
$m^{\prime \prime} \rightarrow 0$ limit where the integral is dominated by the
$v
\sim 0$ region. Simple considerations show that for $T < T_c$
\beq
K(v)\hspace{10pt} {\sim}_{ \hspace{-20pt} \raisebox{-1.ex}{${\scriptstyle m''
\rightarrow 0 }$ } } m^{\prime \prime} v \frac {\Lambda_{\overline
{MS}}}{M_F(T)} \;\;\;,
\eeq
where $M_F(T)$ is given in Eq.~(\ref {gap}). Moreover, $K(v)$ has a cut
starting at a negative value of $v$ and lying along the negative real axis.
Hence, the integral Eq.~(\ref{oint}) converges exponentially to the expected
result as $m^{\prime \prime}
\rightarrow 0$. As $T \rightarrow T_c$, the branching point approaches the
value $v=0$ merging to it at $T=T_c$. At this point the
integral becomes divergent at $m^{\prime \prime}=0$ and does not allow more
extrema for $T > T_c$. Numerical results obtained at different temperatures
(see Fig. 1) indicate that  $m^{\prime \prime}=0$ is in fact the only real
extremum
which leads to
a smooth (second order) phase transition occuring at the critical
temperature
\beq
t_c = 0.57 \,\,\,.
\eeq
The standard large $N$ structure of the phase transition is then exactly
reproduced.

\section{Conclusions}
In this letter, we have shown that a recently proposed variational method can
be
successfully generalized to the finite temperature domain. For comparison
purposes, and to introduce
the method in the study of phase transitions
we have chosen to start with the large $N$ limit of the Gross-Neveu model.
We have seen that optimization problems encountered in the zero temperature low
order
variational calculation persist at finite temperatures. Then, by
applying the all order variational calculation scheme developed in Ref. \cite
{fred} we
were able to recover the usual large $N$ result for the phase transitionof the
GN model, showing that the whole calculational scheme
works remarkably well at all temperatures. The finite $N$ case, which is
interesting due to some important theorems related to phase transitions in one
space dimension is also technically much more complex
and will be discussed in a forthcoming work.

\section{Acknowledgments}
M.B.P. would like to thank CNPq(Brazil) for a post-doctoral grant.

\begin{figure}
\caption{${M_F(T)}/{M_F(0)}$ as a function of the arbitrary parameter
$m^{\prime \prime}$ for different temperatures. From top to bottom the curves
represent $t=0.1$, $t=0.5$ and $t=0.55$ respectively.}
\end{figure}


\begin{references}
\bibitem{pm}P.M. Stevenson, Phys. Rev. {\bf D30},1712(1984); A. Okopi\'nska,
{\it ibid.} {\bf D35},1835(1987); P.M. Stevenson and I. Stancu {\it ibid.}
{\bf D42},2710(1990).
\bibitem{suk} A. Duncan and M. Moshe, Phys. Lett {\bf B215}, 352(1988);
H.F. Jones and M. Moshe, Phys. Lett. {\bf B234},492(1990);
S.K. Gandhi, H.F. Jones and M.B. Pinto, Nucl. Phys. {\bf B359},429(1991);S.K.
Gandhi and M.B. Pinto, Phys. Rev. {\bf D49},4258(1994).
\bibitem{an} A. Neveu, Nucl. Phys. {\bf B18}B(Supp.),242(1990)
\bibitem{gn} D. Gross and A. Neveu, Phys. Rev. {\bf D10},3235(1974)
\bibitem{fred}C. Arvanitis, F. Geniet and A. Neveu, Montpellier preprint PM
94-19 (hep-th/9506188)
\bibitem{fred1}C. Arvanitis, F. Geniet, M. Iacomi, J.-L. Kneur and A. Neveu,
Montpellier preprint PM 94-20
\bibitem{mar}M. B. Pinto, Phys. Rev. {\bf D50},7673(1994)
\bibitem{gnft}L. Jacobs, Phys. Rev. {\bf D10},3956(1974)
\bibitem{hay} B.J. Harrington and A. Yildiz, Phys. Rev. {\bf D11},779(1974)
\bibitem{landau} L.D. Landau and E.M. Lifshtiz, {\it Statistical Physics}
(Pergamon, N.Y., 1958) p.482; S. Coleman, Commun. Math. Phys. {\bf
31},259(1973);N.D. Mermin and H. Wagner, Phys. Rev. Lett. {\bf 17},1133(1966);
R.F. Dashen, S.-K. Ma and R. Rajaraman, Phys. Rev {\bf D11},1499(1975).
\bibitem{witten} E. Witten, Nucl. Phys. {\bf B145},110(1978).
\bibitem{bere} V.L. Berezinski, Sov. Phys. Jetp {\bf 32}, 493(1970); J.M.
Kosterlitz and D.J. Thouless, J. Phys. {\bf C6},1181(1973).
\bibitem{abdalla} E. Abdalla, B. Berg and P. Weiz, Nucl. Phys. {\bf
B157},387(1979); R. K\"{o}berle, V. Kurak and J.A. Swieca, Phys. Rev {\bf
D20},897(1979), Erratum 2638.
\bibitem{davis} A. D'Adda and A.C. Davis, Phys. Lett. {\bf B101},85(1981).
\bibitem{karowski} M. Karawski and M.J. Thun, Nucl. Phys. {\bf B130},224(1977);
E. Witten, Nucl. Phys. {\bf B142},285(1978); R. Shankar and E. Witten, Nucl.
Phys. {\bf B141}349,(1978); M. Karowski and M.J. Thun, Nucl. Phys. {\bf
B190},67(1981).
\bibitem{beno} B. Bellet, P. Garcia, F. Geniet and M.B. Pinto; in preparation
\bibitem{ftemp} D. Bailin and A. Love, {\it Introduction to gauge field theory}
(Adam Higler, Bristol, 1986); R. J. Rivers, {\it Path Integral Methods in
Quantum Field Theories}(CUP, Cambridge,1987)
\bibitem{pms}P.M. Stevenson, Phys. Rev. {\bf D23},2916(1981)
\bibitem{phi}B. Bellet, P. Garcia and A. Neveu Montpellier preprints PM 94-21
(hep-th/9507155) and PM 94-22 (hep-th/9507156)
\end{references}
\end{document}